\documentclass{aa}

\usepackage{txfonts}
\usepackage[english]{babel}
\usepackage{amsmath,amssymb}
\usepackage{graphicx, subfig}
\usepackage[normalem]{ulem}
\usepackage{verbatim}
\usepackage{multirow}
\usepackage{mathtools}
\usepackage{epstopdf}
\usepackage{url}
\usepackage{xcolor,color}
\usepackage{orcidlink}
\usepackage{natbib,twoopt}
\usepackage[hyphenbreaks]{breakurl}
\usepackage{booktabs}
\usepackage{placeins}

\bibpunct{(}{)}{;}{a}{}{,}
\definecolor{cobalt}{rgb}{0.06, 0.2, 0.65}
\hypersetup{
  colorlinks,
  citecolor=cobalt,
  linkcolor=[rgb]{0.8, 0.2, 1.0},
  urlcolor=cobalt,
}
\makeatletter
  \newcommandtwoopt{\citeads}[3][][]{\href{http://adsabs.harvard.edu/abs/#3}%
    {\def\hyper@linkstart##1##2{}%
     \let\hyper@linkend\@empty\citealp[#1][#2]{#3}}}
  \newcommandtwoopt{\citepads}[3][][]{\href{http://adsabs.harvard.edu/abs/#3}%
    {\def\hyper@linkstart##1##2{}%
     \let\hyper@linkend\@empty\citep[#1][#2]{#3}}}
  \newcommandtwoopt{\citetads}[3][][]{\href{http://adsabs.harvard.edu/abs/#3}%
    {\def\hyper@linkstart##1##2{}%
     \let\hyper@linkend\@empty\citet[#1][#2]{#3}}}
  \newcommandtwoopt{\citeyearads}[3][][]%
    {\href{http://adsabs.harvard.edu/abs/#3}
    {\def\hyper@linkstart##1##2{}%
     \let\hyper@linkend\@empty\citeyear[#1][#2]{#3}}}
\makeatother

\def\apjl{ApJL}
\def\apjs{ApJS}
\def\aap{A\&A}

\def\be{\begin{equation}} 
\def\ee{\end{equation}} 
\def\ba{\begin{eqnarray}} 
\def\ea{\end{eqnarray}}

\def\kms{\,{\rm {km\, s^{-1}}}} 
 
\def\msun{{\Msun}}

  \def\cm{${\rm cm}$}

\def\gsim{\lower.5ex\hbox{\gtsima}} 
\def\lsim{\lower.5ex\hbox{\ltsima}} \def\gtsima{$\; \buildrel > \over 
\sim \;$} \def\ltsima{$\; \buildrel < \over \sim \;$} \def\prosima{$\; 
\buildrel \propto \over \sim \;$} \def\gsim{\lower.5ex\hbox{\gtsima}} 
\def\lsim{\lower.5ex\hbox{\ltsima}} 
\def\simgt{\lower.5ex\hbox{\gtsima}} 
\def\simlt{\lower.5ex\hbox{\ltsima}} 
\def\simpr{\lower.5ex\hbox{\prosima}}   
  
 \def\gtsima{$\; \buildrel > \over \sim \;$} 
\def\ltsima{$\; \buildrel < \over \sim \;$} 
\def\gsim{\lower.5ex\hbox{\gtsima}} 
\def\lsim{\lower.5ex\hbox{\ltsima}} 
\def\simgt{\lower.5ex\hbox{\gtsima}} 
\def\simlt{\lower.5ex\hbox{\ltsima}} 
\def\simpr{\lower.5ex\hbox{\prosima}}

\def\msun{\,{\rm \Msun}}

\def\E3{{\cal E}_{\rm g}^{III}}

\def\Msun{\rm M_\odot}
\def\Zsun{\rm Z_\odot}
\def\Lsun{{\rm L}_\odot}

\def\r12{r_{1/2}} 
\def\x12{x_{1/2}}

\def\cm2g{ {\rm cm^2} \, {\rm {g^{-1}}}}

\newcommand\code[1]{\textsc{\MakeLowercase{#1}}}
\newcommand{\quotes}[1]{``#1''}

\def\nh2{n_{\rm H2}}
\def\fh2{f_{\rm H2}}

\def\angstrom{\textrm{A\kern -1.3ex\raisebox{0.6ex}{$^\circ$}}}

\def\msun{{\rm M}_{\odot}}

\def\TdMWS20{${\langle T_{\rm d} \rangle}_{\rm M}^{\rm S20}$}
\def\TdLWS20{${\langle T_{\rm d} \rangle}_{\rm L}^{\rm S20}$}

\usepackage{siunitx}
\usepackage{cleveref}
\usepackage{xspace}

\crefformat{equation}{eq.~#2#1#3}
\crefformat{figure}{Fig.~#2#1#3}
\crefformat{section}{Sec.~#2#1#3}
\crefformat{subsection}{Sec.~#2#1#3}
\crefformat{table}{Tab.~#2#1#3}

\newcommand{\dd}{\mathop{\mathrm{d}\!}{}}
\newcommand{\deriv}[2]{\dfrac{\dd #1}{\dd #2}}

\newcommand{\ped}[1]{_{\rm #1}}

\def\muv{{\rm M_{\rm UV}}}

\newcommand{\fpbh}{f\ped{PBH}}

\newcommand{\npbh}{n\ped{PBH}}
\newcommand{\Mpbh}{M\ped{PBH}}

\defcitealias{matteri:2025}{M25}

\graphicspath{{plots/}}

\linenumbers
\renewcommand\makeLineNumber{}

\begin{document}

\title{Beyond the first galaxies primordial black holes shine}
\titlerunning{Beyond the first galaxy PBHs shine}

\author{
Antonio Matteri$^{1}$ \orcidlink{0009-0007-9985-9112} \and
Andrea Ferrara$^{1}$ \orcidlink{0000-0002-9400-7312}\and 
Andrea Pallottini$^{2}$ \orcidlink{0000-0002-7129-5761} 
       }
\authorrunning{Matteri et al.}
\institute{
Scuola Normale Superiore, Piazza dei Cavalieri 7, 56126 Pisa, Italy
\and Dipartimento di Fisica ``Enrico Fermi'', Universit\`{a} di Pisa, Largo Bruno Pontecorvo 3, Pisa I-56127, Italy
}
\date{Received: March 24, 2025; Accepted: July 14, 2025}

\abstract
{
The presence of nine candidate galaxies at $z=17$ and $z=25$ discovered by the \textit{James Webb Space Telescope} in relatively small sky areas, if confirmed, is virtually impossible to reconcile with the predictions of the current galaxy formation model. We show here that the implied UV luminosity density can be produced by a population of primordial black holes (PBHs) of mass $\Mpbh = 10^{4-5} \, \Msun$ residing in low-mass halos ($M_h \approx 10^{7} \, \Msun$), and accreting at a moderate fraction of the Eddington luminosity, $\lambda_E \simeq 0.36$. These sources precede the first significant episodes of cosmic star formation. At later times, as star formation is ignited, PBH emission becomes comparable to, or subdominant with respect to, the galactic emission. This PBH+galaxy scenario reconciles the evolution of the UV luminosity function (LF) from $z=25$ to $z=11$. If ultra-early sources are powered purely by accretion, this strongly disfavours seed production mechanisms requiring the presence of stars (massive stars, Pop III stars, or clusters), or their UV radiation (direct collapse BHs), leaving PBHs as the only alternative solution available so far. Alternative explanations, such as isolated, large clusters ($\approx 10^7 \,\Msun$) of massive ($m_\star =10^3 \Msun$) Pop III stars are marginally viable, but require extreme and unlikely conditions that can be probed via UV and far-infrared (FIR) emission lines or gravitational waves.
}


\maketitle

\section{Introduction} \label{sec:intro}

Since the beginning of its operations in July 2022, the \textit{James Webb Space Telescope} (JWST) has obtained a number of pivotal results that are revolutionizing our view of the early Universe. Remarkably, JWST observations have led to the discovery \citep{naidu:2022, castellano:2022, atek:2023, labbe:2023} and spectroscopic confirmation \citep{arrabal:2023, bunker:2023, curtis:2023, robertson:2023, wang:2023, hsiao:2024, zavala:2024} of super-early ($z>10$) galaxies up to redshift $z\approx 14$ \citep{carniani:2024}, when the Universe was $\approx 300\,\rm Myr$ old.
These results challenge expectations based on pre-JWST data, theoretical predictions, and perhaps even $\Lambda$CDM-based\footnote{Throughout the paper, we assume a flat Universe with the following cosmological parameters: $\Omega_{m} = 0.3111$, $\Omega_{\Lambda} = 1- \Omega_{m}$, $\Omega_{r} = 9.1\times 10^{-5}$, and $\Omega_{b} = 0.04897$, $h=0.6766$, $\sigma_8=0.8102$, where $\Omega_{m}$, $\Omega_{\Lambda}$, $\Omega_{r}$, and $\Omega_{b}$ are the total matter, vacuum, radiation and baryon densities, in units of the critical density; $h$ is the Hubble constant in units of $100\,\kms \,\rm Mpc^{-1}$, and $\sigma_8$ is the late-time fluctuation amplitude parameter \citep{planck:2018}.} galaxy formation scenarios. 

Super-early galaxies are characterized by bright UV luminosities ($\muv \lesssim -20$), steep UV spectral slopes ($\beta \lesssim -2.2$, \citealt{Topping_2022, Cullen_2024, Morales_2024}), compact sizes (effective radius $r_e \lesssim 100$ pc, \citealt{Baggen_2023, Morishita_2024, carniani:2024}), large (for their epoch) stellar masses ($M_\star \approx 10^9 \,\Msun$), and thus are sometimes nicknamed ``Blue Monsters'' \citep{Ziparo23}. Explaining their relatively large comoving number density ($n \approx 10^{-5}-10^{-6}\ \rm cMpc^{-3}$; see, e.g., \citealt{Casey_2024, Harvey_2024}) requires the introduction of new physical ingredients in standard galaxy evolution models.

These additions include (a) high star formation variability \citep{Furlanetto_22, Mason23, Mirocha23, Sun_2023, Pallottini23, gelli:2024}, (b) reduced feedback resulting in a higher star formation efficiency \citep[feedback free model, FFB,][]{Dekel23, Li23}, (c) a top-heavy IMF or primordial stars \citep{Inayoshi22, Wang_2023, Trinca24, Hutter_2024, Cueto23}, and (d) very low dust attenuation \citep[Attenuation-Free Model, AFM,][]{ferrara:2023, Ziparo23, Fiore23, Ferrara24a, Ferrara24b, ferrara:2024_outflow}. These models have been, to varying degrees, successful in explaining, at least in part, the statistical or evolutionary properties of the newly discovered super-early galaxy population. 

However, JWST data continue to be a source of genuine surprises. \citet{perez:2025} have recently reported the discovery of six F200W and three F277W robust dropout sources that are consistent with being at $z \approx 17$ and $z\approx 25$, respectively \citep[see also][]{harikane:2024, kokorev:2024, whitler:2025, Gandolfi25, castellano:2025}. The detection, if confirmed, implies exceptionally high number densities; i.e., for $\muv=-18$ sources $\approx 10^{-4.2} (10^{-5}) \, \rm Mpc^{-3}$  at $z=17$ ($z=25$) which, yet again, are at odds with all current models. 
These ultra-early galaxies, if confirmed spectroscopically, effectively falsify all the solutions proposed above. We are therefore compelled to explore ``beyond standard model'' (BSM) scenarios to explain their existence in such large numbers. Specifically, this means considering modifications to the $\Lambda$CDM model.

This typically entails assuming different cosmological scenarios that enhance the halo mass function (HMF) at $z>10$ \citep{haslbauer:2022, boylan:2023}, either through effective modification of the transfer function \citep{padmanabhan:2023}, or by exploring different models, such as an early dark energy contribution at $z\approx3400$ \citep{klypin:2021, shen:2024}, the tilt of the primordial power spectrum \citep{parashari:2023}, the importance of non-Gaussianities \citep{biagetti:2023}, and the presence of primordial black holes \citep[PBHs,][]{liu:2022, matteri:2025}.

Among these modifications, solving the overabundance problem with PBHs is appealing, since their impact tends to decrease with cosmic time. For this reason, PBH solutions do not cause
insurmountable tensions at $z<10$, contrary to other BSM proposals \citep{gouttenoire:2023,sabti:2024}.
This solution is also particularly attractive as it is known that some of the super-early galaxies (GN-z11, \citealt{Maiolino23b}; GHZ2, \citealt{castellano:2024}; GHZ9, \citealt{Napolitano24}; UHZ1, \citealt{natarajan:2024}) contain massive black holes that contribute significantly to the observed luminosity of the source.
In addition, PBHs could provide optimal seeds from which supermassive BHs can subsequently grow \citep{rubin:2001, duchting:2004, khlopov:2010, kawasaki:2012, dayal:2024, ziparo:2024}. 

In this paper, building on the results from \citet[][\citetalias{matteri:2025} hereafter]{matteri:2025}, we propose that the very slow measured decrease of $\rho_{\rm UV}(z)$ at $z>15$ is sustained by PBHs even before the first galaxies form in detectable numbers. Hence, ultra-early galaxies could be systems whose emission is completely dominated by accretion onto PBHs, rather than of stellar origin.

\section{Model}\label{sec:model}

\citetalias{matteri:2025} presented an effective model, calibrated using data at lower redshifts, to predict the luminosity function (LF) of super-early galaxies, including the additional PBH contribution. \citetalias{matteri:2025} show that such a model can physically explain the overabundance of bright galaxies observed at $z= 11$.
Primordial black holes can affect the LF in two distinct ways: by (a) enhancing the dark matter power spectrum due to their discrete spatial distribution, and/or (b) contributing accretion luminosity to the host galaxy, i.e., as an Active Galactic Nucleus (AGN).

The distribution of PBHs at formation is modelled by a lognormal mass function with characteristic mass $\Mpbh$, standard deviation $\sigma$, and accounting for a mass fraction $\fpbh$ of the dark matter content of the Universe. Their luminosity is produced at a constant Eddington ratio $\lambda_E = L/L_E$, where $L_E$ is the Eddington luminosity. The original model does not account for the growth of PBHs with time (that is, the PBH mass is assumed to be constant). Stellar emission is accounted for by an effective astrophysical model whose parameters are calibrated on the measured LFs in the range $4 < z < 9$ (see \citetalias{matteri:2025}, Fig. 2).

Within these simplified settings, the overabundance of bright galaxies at $z \approx 11$ is explained by the model when a small fraction ($\fpbh\approx10^{-8}$) of low-mass ($\Mpbh\approx10^4\,\Msun$) PBHs accrete at super-Eddington ($\lambda_E\approx10$) rates, without violating any current constraint from the cosmic microwave background \citep[CMB,][]{nakama:2018, serpico:2020}.

In the present work, we extend the \citetalias{matteri:2025} model by considering PBH growth self-consistently with accretion as follows. For a given radiative efficiency, $\eta$, we assumed that a PBH formed at the radiation-matter equality redshift, $z_{\rm eq}=3400$ \citep{planck:2018}, with mass $M_0$ by redshift $z < z_{\rm eq}$ will have a mass
\begin{equation}
    \label{eq:mass_evolution}
    M(z) = M_0 \exp\Bigg[\frac{t(z)-t(z_{\rm eq})}{t_{S}}\lambda_E\frac{1-\eta}{\eta}\Bigg] \equiv M_0 g(z, z_{\rm eq})
\end{equation}
where $t_{S} = \SI{450}{Myr}$ is the Salpeter time \citep{salpeter:1964}, $t(z)$ is the age of the Universe at redshift $z$, and $g(z, z_{\rm eq})$ is the implicitly defined growth function. We fixed $\eta=0.1$ and left $\lambda_E$ as a parameter to be determined by matching the data.

Evolving the initial lognormal distribution via \cref{eq:mass_evolution} retains this functional form once the parameters are updated as follows:
\begin{align}
\label{eq:lognormal_evolution}
[\Mpbh, \fpbh](z) &= g(z,z_{\rm eq}) [\Mpbh, \fpbh](z_{\rm eq})\\
\sigma(z) &= \sigma(z_{\rm eq}) \nonumber.
\end{align}
This growth shifts the logarithm of the PBH mass by a constant, affecting only the peak position and leaving the shape of the distribution unaltered. Relaxing the approximation of a constant and uniform $\lambda_E$ would lead to a more complex evolution, i.e, we expect the scatter to increase. This means that either a different functional form for the mass function is produced at late times or that at least $\sigma$ grows with cosmic time. Unfortunately, current data quality prevents us from adopting a more refined framework.

Following \citetalias{matteri:2025}, we seeded PBHs in halos above a redshift-dependent mass $M_{min}$, implicitly defined by  
\begin{equation}\label{eq:Mmin_definition}
    \int_{M_{min}(z)}^{\infty} \deriv{n}{M_h}(M_h, z) \dd M_h = n_0 \approx \SI{7.13}{Mpc^{-3}}\,,
\end{equation}
where $n_0$ is the comoving density of halos corresponding to $M_{min} =10^{7.5}\,\Msun$ at $z=11$ and $\dd n/\dd M_h$ is the HMF.
Although this choice can be justified by the physical arguments discussed in \citetalias{matteri:2025}, it is admittedly uncertain. Nevertheless, we have checked that it has only a very minor impact on our results as long as $M_{min}$ is roughly $5\times 10^7\,\Msun$ at $z=11$.
We note that in our scheme, both PBHs and dark matter halos grow with time, but at different rates, respectively given by \cref{eq:lognormal_evolution} and the cosmological accretion rate governing the HMF evolution (\cref{eq:Mmin_definition}).

We note that for any ($\sigma$, $\lambda_E$) pair, a degeneracy exists between the other two parameters, $\Mpbh$, and $\fpbh$, such that the derived UV LF depends only on the ratio of these two quantities. These must satisfy, at $z=11$, the relation $\Mpbh \fpbh^{-1} \approx 10^{11.7}\, \msun$ \citepalias{matteri:2025}.    
Applying \cref{eq:lognormal_evolution}, we next constructed the predicted UV LF at any redshift. 

Finally, from the UV LF, we computed the UV luminosity density, $\rho_{\rm UV}$, which is defined as
\begin{equation}\label{eq:rhouv_definition}
    \rho_{\rm UV}(z) = \int^{M_{\rm UV}^{\rm lim}}_{-\infty}L_{\rm UV}(\muv) \Phi_{\rm UV}(\muv, z) \dd \muv
\end{equation}
where $L_{\rm UV}$ is the specific UV luminosity corresponding to the magnitude $\muv$, $\Phi_{\rm UV}$ is the LF, and $\rm M_{\rm UV}^{\rm lim}$ is the faint-end magnitude limit, conventionally set to $-17$, as in the observational data adopted below.

Similarly to \citetalias{matteri:2025}, we inferred the best parameters of this model in a Bayesian framework by running
a Monte Carlo Markov Chain (MCMC). We used $\fpbh$, $\Mpbh$, $\sigma$, and $\lambda_E$ as parameters and the luminosity function
data at $z=11$ \citep[from ][]{donnan:2024, mcleod:2024}, $17$, and $25$ \citep[both from][]{perez:2025} as constraints.
Flat priors were used on the parameters over the intervals listed in \cref{tab:priors}. Limits on $\Mpbh$ were chosen
to avoid both small PBHs and violation of current limits from CMB distortions \citep{byrnes:2024, pritchard:2025}.

\begin{table}[ht]
    \centering
    \caption{Prior limits on MCMC parameters}
    \begin{tabular}{c | c c}
        \toprule
            Parameter & Low & High \\
        \midrule
            $\log \fpbh$         & -10  & 0   \\
            $\log \Mpbh / \Msun$ & 0    & 5   \\
            $\sigma$             & 0.01 & 1.5 \\
            $\log \lambda_E$          & -2 & 0.5 \\
        \bottomrule
    \end{tabular}
    \label{tab:priors}
\end{table}

\section{Results}\label{sec:results}

\begin{figure}[h]
    \centering
    \includegraphics[width=0.90\linewidth]{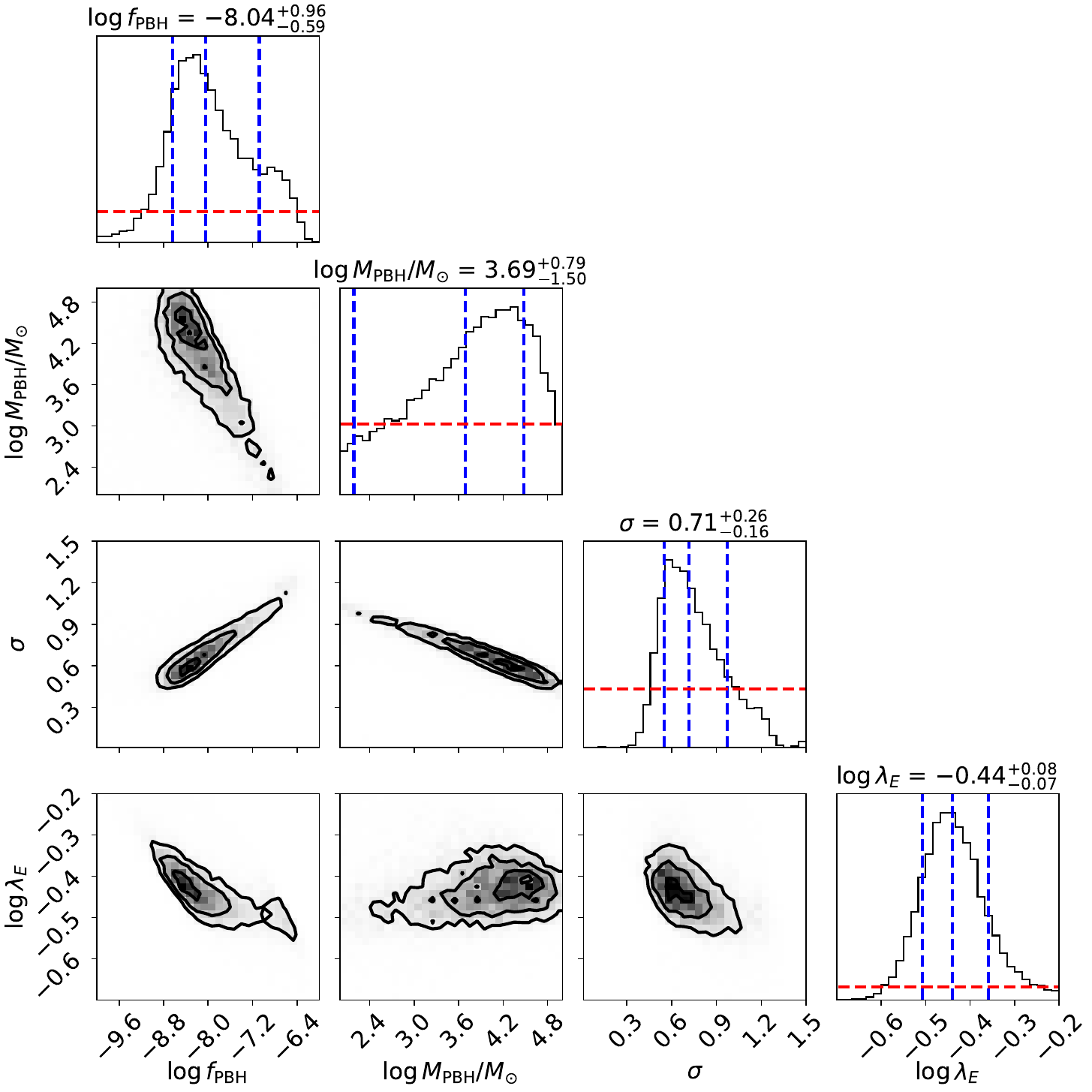}
    \caption{
        Corner plot of MCMC results. Priors on parameters are
        constant over the intervals reported in \cref{tab:priors} and
        are plotted as dashed red lines on the 1D distributions.
        The parameters' median, 16th, and 84th percentiles are indicated both
        above the 1D distributions and as vertical dashed blue lines. 
    }
    \label{fig:corner-plot}
\end{figure}

\begin{table*}
    \centering
    \caption{Fiducial values of parameters regulating PBH properties and evolution}
    \begin{tabular}{c | c c c c c }
         \toprule
         $z$ & $\log \fpbh (z)$ & $\log \Mpbh(z)\, [\Msun]$ & $\sigma$ & $\log \lambda_E$ & $\log M_{min}\,[\Msun]$ \\
         \midrule
         $z_{\rm eq}=3400$ & $-8.02^{+0.97}_{-0.63}$ & $3.63^{+0.81}_{-1.42}$ & $0.72^{+0.25}_{-0.16}$ & -- & -- \\
            25 & $-7.62^{+0.95}_{-0.54}$ & $4.08^{+0.79}_{-1.44}$ & $0.72^{+0.25}_{-0.16}$ & $-0.44^{+0.08}_{-0.07}$ & $5.78$\\
            17 & $-7.31^{+0.92}_{-0.49}$ & $4.39^{+0.80}_{-1.45}$ & $0.72^{+0.25}_{-0.16}$ & $-0.44^{+0.08}_{-0.07}$ & $6.76$\\
            11 & $-6.71^{+0.86}_{-0.41}$ & $5.03^{+0.81}_{-1.49}$ & $0.72^{+0.25}_{-0.16}$ & $-0.44^{+0.08}_{-0.07}$ & $7.50$\\
         \bottomrule
    \end{tabular}
    \tablefoot{
    The PBHs are distributed with a lognormal mass distribution with peak $\Mpbh$ with constant s.d. $\sigma$ (see \cref{eq:lognormal_evolution}). They grow via accretion according to \cref{eq:lognormal_evolution} with constant Eddington ratio $\lambda_E$.
    \label{tab:results}
    }
\end{table*}

\begin{figure*}
    \centering
    \includegraphics[width=0.49\linewidth]{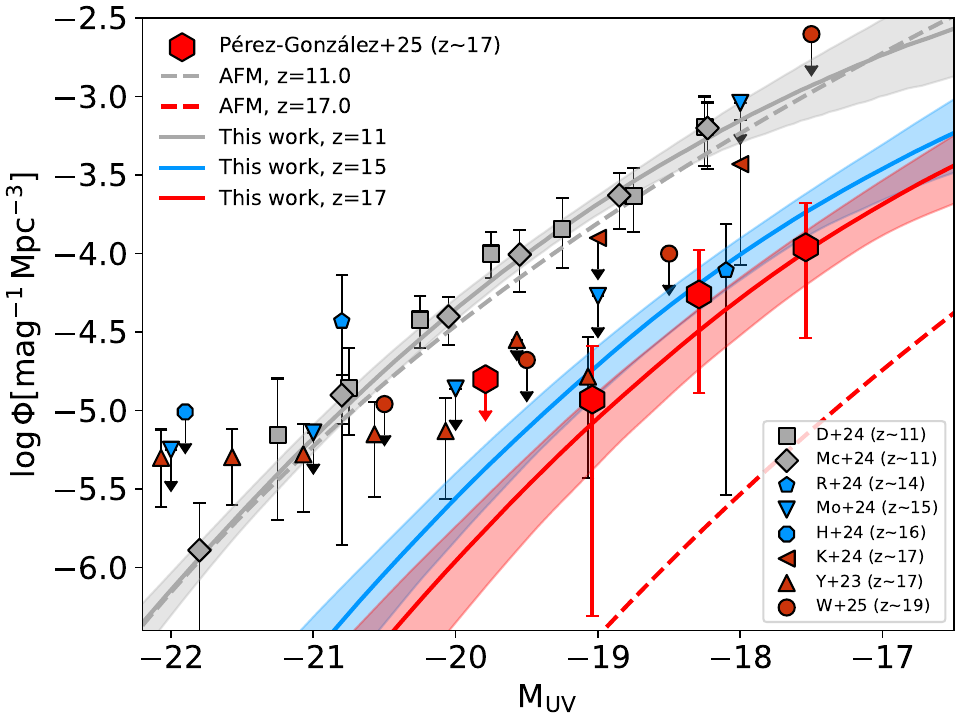}
    \includegraphics[width=0.49\linewidth]{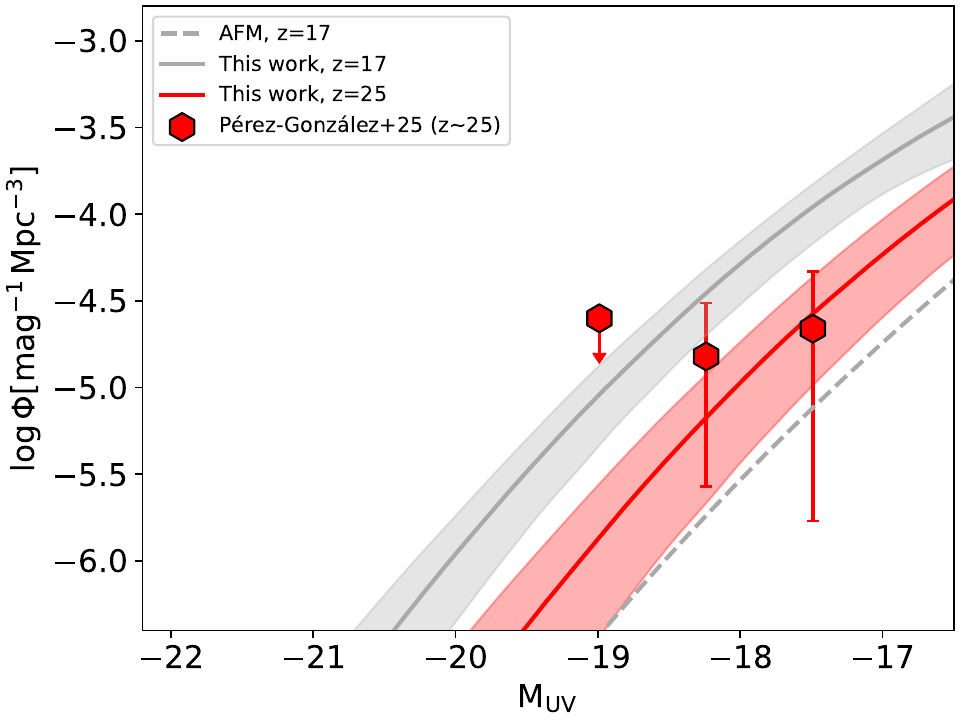}
    \caption{
        UV luminosity functions at redshift $z=11$, $z=15$, $z=17$ (left and right panels), and $z=25$ (right panel). Solid lines represent our median predictions (galaxies plus PBH emission) with credible intervals at the 16th and 84th percentiles shown as shaded areas. Dashed lines show the AFM results \citep{ferrara:2023} for galaxies only. Red hexagons denote $z=17$ and $25$ observations from \citet{perez:2025}; other points represent data in the range $11 \le z \le 19$ from \citet{donnan:2024, mcleod:2024, robertson:2024, moroshita:2024, harikane:2024, kokorev:2024, yan:2023, whitler:2025}.
        \label{fig:fig01}
    }
\end{figure*}

\begin{figure}
    \centering
    \includegraphics[width=0.99\linewidth]{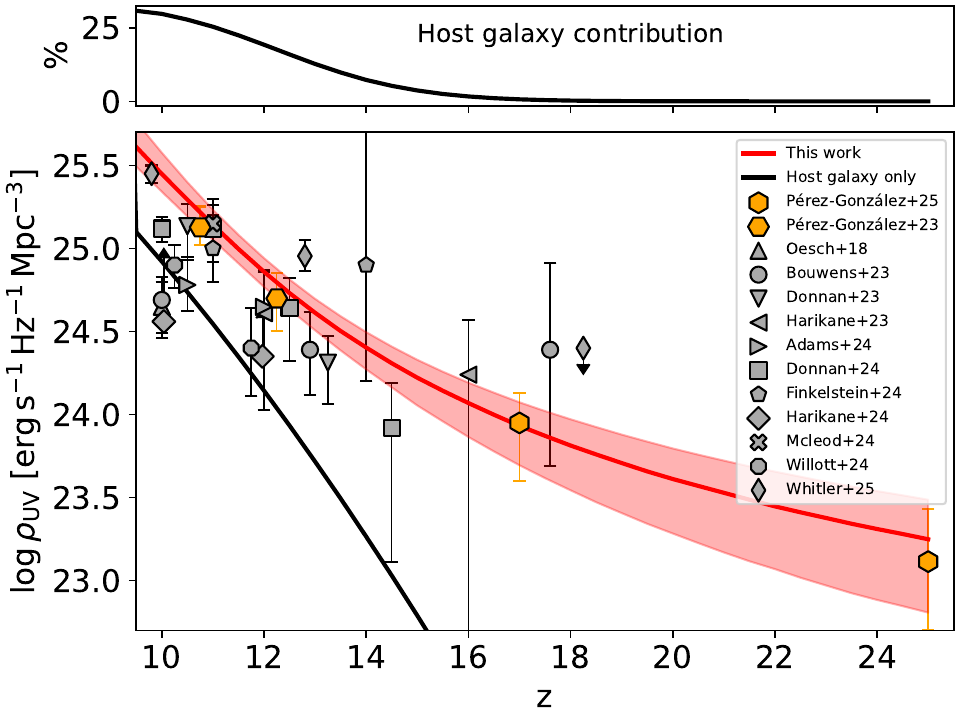}
    \caption{
        UV luminosity density (integrated to a limiting magnitude ${\rm M}_{\rm UV}^{\rm lim}=-17$) evolution with redshift.
        Our fiducial galaxy plus PBH model is shown by the red curve and the galaxy-only model by the black curve. Shaded areas indicate credible intervals between 16th and 84th percentiles. Orange hexagons represent observational data from \citet{perez:2023, perez:2025}; gray symbols are from \citet{Oesch18, bouwens:2023, donnan:2023, harikane:2023, adams:2024, donnan:2024, finkelstein:2024, harikane:2024, mcleod:2024, Willott24, whitler:2025}. \textit{Top panel:}  Fractional UV luminosity contribution of the host galaxy relative to the total.
        \label{fig:fig02}
    }
\end{figure}

The MCMC analysis yields a one-peaked parameter distribution (see corner plot reported in \cref{fig:corner-plot}),
which we proceed to sample to build a fiducial model. 
In \cref{fig:fig01} we present the UV LFs predicted by our fiducial model at different redshifts (see \cref{tab:results} for the parameters at each redshift of interest).
At $z=11$, the model matches the observed data well. At $M_{\rm UV} \simeq -21 (-19)$, PBHs contribute about 75\% (50\%) of the UV luminosity of the source, with the remaining 25\% (50\%) produced by stars. Notably, at $z=11$ our model is virtually indistinguishable from the AFM (dashed line), which includes only unattenuated radiation from stars (i.e., no PBHs). 

At $z=17$ the AFM predicts a $\approx 3$ dex UV LF drop at $\muv = -19$, driven essentially by the steep decline of the HMF, resulting in a source comoving density $\lesssim 10\times$ the one observed by \citet{perez:2025}, if confirmed. In contrast, our model fits the data well, as PBH emission, regulated by mass growth, declines less steeply than the HMF. Consequently, the observed luminosity becomes largely ($> 90\%$) dominated by PBHs.

This trend persists as we extend our analysis to $z=25$, as illustrated in the right panel of Fig. \ref{fig:fig01}. Remarkably, the observed UV LF, if confirmed, shows only a relatively modest tenfold decrease, surpassing even the $z=17$ predictions (which align well with data up to $z=11-12$) of galaxy formation models such as the AFM \citep{ferrara:2024_outflow}, feedback-free models \cite{Li23}, and semi-analytical approaches \citep{somerville:2025}. It should be noted that even if only one of the candidates is confirmed, the strong tension with the previous models remains unresolved. This suggests that standard galaxy formation models lack a crucial component, which we propose to be very early or primordial black holes. Indeed, when these are incorporated, our model consistently reproduces the $z=25$ UV LF (red curve).

We corroborate these conclusions by examining the predicted redshift evolution of the UV luminosity density, $\rho_{\rm UV}$, compared to available data in \cref{fig:fig02}. The black line represents our model without PBH accretion, where UV light is produced solely by stars. Consistent with most galaxy evolution models, $\rho_{\rm UV}$ is in good agreement with the data at $z\approx 10$, but declines dramatically at early epochs. Including PBH emission, however, recovers the observed data, if confirmed, producing a slower decline in $\rho_{\rm UV}$.

This feature can be understood by first noting (upper panel in \cref{fig:fig02}) that the UV contribution from stars in the host galaxy drops below 10\% by $z=14$, indicating that ultra-early sources are dominated by PBH light. Combining this result with the relatively slow decrease of the PBH peak mass from $\log (\Mpbh/\Msun) = 5.03^{+0.81}_{-1.49}$ at $z=11$ to $\log (\Mpbh/\Msun) = 4.08^{+0.79}_{-1.44}$ at $z=25$, driven by a moderate accretion rate ($\lambda_E \approx 10^{-0.44} \approx 0.36$), provides the final interpretation of the trend. Thus, light production in the Universe shifts from stars to accretion onto PBHs towards cosmic dawn.

A magnitude $\muv \approx -18$ corresponds to a UV luminosity produced by a PBH with mass $2.2\times 10^6\,\Msun$ accreting at $\lambda_E=0.36$, assuming a bolometric correction $f_{\rm bol}(L_{\rm UV})=0.13$ \citep{shen:2020}. This represents a relatively rare ($\approx 3 \sigma$) peak in the PBH mass distribution. The corresponding accreting PBH X-ray luminosity is well below the current \textit{Chandra} detection limits \citep{Xue16}.

The comoving number density of PBHs is $\npbh \propto \fpbh/\Mpbh$ when $\sigma$ is fixed. Such a ratio, according to \cref{eq:lognormal_evolution} (see also eq. 8 in \citetalias{matteri:2025}), does not evolve with redshift.
This arises because PBHs only form during the radiation-dominated era and are assumed to not evaporate or merge later on. As a result, the model parameters reported in \cref{tab:results} yield $\npbh = 0.013^{+1.262}_{-0.012} \, \si{Mpc^{-3}}$. Using the result in \cref{eq:Mmin_definition}, this implies that only a small fraction ($\npbh/n_0 \approx 10^{-3}$ at median density) of halos above $M_{min}$ actually host PBHs. 
We can use this argument to calculate the typical mass of halos hosting a PBH shining at $\muv = -18$ at $z=25$ to be $\log (M_h/\Msun) \approx 7.5$, corresponding to a ratio $\Mpbh/M_h = 0.07$.

\section{Discussion}

The result above indicate that ultra-high ($z>15$) redshift galaxies, if confirmed, pose an exceptionally difficult challenge to theoretical models of galaxy formation. Motivated by this urgency, we have explored an alternative solution provided by PBHs. We find that ultra-early galaxies must be largely, if not entirely, dominated by emission from accretion onto intermediate-mass PBHs rather than by stars. 

This scenario is sufficiently extreme to entail a number of implications that warrant a closer inspection. Here, we limit our discussion to the most obvious ones. However, before doing so, we note that if the $z=25$ candidates are powered by PBHs, we expect them to be very compact and show an almost point-like morphology. Visual inspection of the cutout images of the three objects appears to confirm this hypothesis. 

Additionally, if we are observing the light from the accretion disk, the UV continuum spectrum should follow a simple power law of the form $F_\lambda \propto \lambda^\beta$ with $\beta = -2.33$ \citep{Shakura73}. The values measured for the three most distant objects in the \cite{perez:2023} sample (midis-z25-1, midis-z25-2, and midis-z25-3) are $\beta = (-2.6 \pm 0.4, -3.0\pm 0.4, -2.4\pm 0.6)$. These are broadly consistent within 1$\sigma$, although we note that these values were derived from only two photometric bands and therefore are likely to be highly uncertain.

The abundance of PBHs is constrained by several observables that would be affected by their presence \citep{carr-kohri:2021}. These translate into upper limits on the PBH abundance as a function of mass.
For masses on the order of $10^{2-5}\, \Msun$, these constraints tighten with $\Mpbh$, and PBHs more massive than $10^5\,\Msun$ are virtually incompatible with CMB observations due to the production of an undetected $\mu$-distortion \citep{chluba:2012,nakama:2018,sharma:2024}. Our model requires a relatively small ($\fpbh\approx 10^{-8}$) population of mid-mass ($\Mpbh\approx10^{4}\,\Msun$) PBHs (\cref{tab:results}), whose parameters $(\Mpbh, \fpbh)$ lie within the current constraints assuming the monochromatic approximation, i.e., assuming all PBHs have the same mass $\Mpbh$. However, the long tail of the lognormal produces a small amount ($2\sigma$ outliers) of $\gtrsim 10^{5}\,\Msun$ PBHs, which are in tension with these constraints. As stated in the literature \citep{nakama:2018,byrnes:2024,pritchard:2025}, monochromatic limits relax when primordial fluctuations are assumed to be non-Gaussian and are also affected by the specific shape of the PBH mass function \citep{wang:2025}, implying that the typical PBHs in our model may not be excluded. That is, a mass function with a shorter tail could comply with current limits, while still explaining observations at $z=25$ as the most massive outliers, eventually requiring a faster accretion onto PBHs or mergers. We defer exploration of this scenario to future work beyond this paper.

\subsection{Implications of the PBH scenario}

\paragraph{Nature of seeds} The first implication of our results is that beyond the first galaxies, PBHs were already active and shining. If ultra-early sources are powered by accretion onto BHs, then a stellar origin for the seeds becomes unlikely. 

Two popular pathways for seed formation \citep[for a review, see][]{Latif16}, core-collapse supernovae and dense stellar clusters, both require the presence of stellar progenitors. The third pathway, direct collapse black holes, requires a strong UV background to prevent fragmentation caused by H$_2$ formation; this UV radiation must also be initially produced by stars. 

The PBHs required by our model have characteristic masses of  $10^{4-5}\, \Msun$ already at $z=25-17$. Therefore, they could serve very well as supermassive black hole seeds. Indeed, considering their growth, their mass approaches those inferred for GN-z11 at $z=10.6$, i.e., $\log (M_\bullet/\Msun) = 6.2 \pm 0.3$ \citep{Maiolino24N}.

\noindent\paragraph{Reionization} PBHs could begin to emit radiation as early as $z \approx z_{\rm eq}$, potentially illuminating the cosmic Dark Ages ($60<z<1100$), the epoch before the formation of the first stars. This emission could, in effect, \quotes{erase} the Dark Ages by providing an early source of light.

Additionally, PBHs may influence the reionization of cosmic hydrogen. The comoving ionization rate density is given by the relation $\dot N_{\rm ion} = \xi_{\rm ion} f_{\rm esc} \rho_{\rm UV}$, with the 1500\angstrom\ -- LyC conversion factor $\xi_{\rm ion} \approx 10^{26}\ \rm Hz\ erg^{-1}$ appropriate for accreting black holes \citep{Madau99}. This yields $\dot N_{\rm ion} = 10^{49} f_{\rm esc}\ \rm s^{-1}\ Mpc^{-3}$ at $z=25$ and $\dot N_{\rm ion} = 10^{51} f_{\rm esc}\ \rm s^{-1}\ Mpc^{-3}$ at $z=12$. Photoionization is balanced by recombinations in the intergalactic medium, giving an equilibrium cosmic hydrogen fraction $(1-x_{\rm HI}) \approx \dot N_{\rm ion} t_{\rm rec}/n_H \approx 0.002 f_{\rm esc}$ at $z=25$ and $(1-x_{\rm HI}) \approx 0.2 f_{\rm esc}$ at $z=12$. Hence, ionization due to PBHs is expected to be very modest, even if $f_{\rm esc}=1$. These values could also remain undetected in the Thomson optical depth of the CMB, considering that a reasonable escape fraction of $f_{\rm esc} \approx 0.1$ would yield a reionization epoch very similar to that expected from observations \citep{greig:2017}.
However, it is likely that, in the configuration where the PBH accretes at the center of a $M_h \approx 10^7 \,\Msun$ halo at sub-Eddington rates (and thus without radiation-driven outflows clearing the gas), $f_{\rm esc} \ll 1$, given the typical absorbing gas column density ($\gtrsim 10^{21}\ \rm cm^{-2}$) in these primordial objects.  

Such PBHs could also impact reionization by heating the intergalactic medium (IGM) via X-ray emission. Although the very dense Universe in which these objects accrete may prevent X-rays from escaping, that is, the accreting gas may become Compton thick, we note that the abundance of these objects is very low. \citet{ricotti:2004} consider a scenario in which PBHs pre-ionize the IGM at $z \gtrsim 20$, showing that a fraction $\approx6.8\times 10^{-6}$ of baryons must be accreted to significantly impact the medium, i.e., ionize about $50\%$ of the hydrogen. In our model, PBHs constitute a fraction $\fpbh\approx 2\times10^{-7}$ of dark matter at $z=11$, implying that they accrete up to $\fpbh\Omega_{dm}/\Omega_{b}\approx 10^{-6}$ of the total baryonic mass by that time, which is still below their limit.

\subsection{Could they be Pop III stars?}

As an alternative to the PBH hypothesis, we can speculate that the observed sources are powered by Pop III stellar clusters.
The (bolometric) light-to-mass ratio of individual massive Pop III stars increases from $\Upsilon=1405\, \Lsun/\Msun$ for a $m_\star =15\, \Msun$ star to $\Upsilon=2.7\times 10^4 \,\Lsun/\Msun$ at $m_\star = 10^3 \Msun$ \citep{scaherer:2002, raiter:2010}. To maximize detectability, we assume that a cluster of $m_\star= 10^3 \Msun$ stars has already formed at $z=25$.
As a UV bolometric correction in the rest-frame wavelength range $1300-1600\ \angstrom\,$, approximately covered by the NIRCam F356W filter, we obtain $f_{\rm bol} \simeq 0.016$, appropriate for a black-body spectrum with $\log (T_{\rm eff}/{\rm K}) = 5.026$. 
A magnitude $\muv= -18$, as observed for the two faintest sources at $z=25$, if confirmed, translates to a UV luminosity $L_{\rm UV} = 10^{9.55} \,\Lsun$. Combining these results, we conclude that a total Pop III stellar mass of $ M_{\star, \rm III} \approx 8.2\times 10^6 \,\Msun$ is required, equivalent to roughly $8200$ stars each with $m_\star=10^3\, \Msun$. 

These massive stars have very short lifetimes, $t_\star = \SI{1.9}{Myr}$ \citep{scaherer:2002}, yielding an approximate duty cycle $f_{\rm duty} = t_\star/t(z=25)= 0.014$, where $t(z=25)=133.2$ Myr is the cosmic age at $z=25$. Given the observed comoving number density ($n \approx 10^{-5}\, \rm Mpc^{-3}$) of such objects and the duty cycle, the cluster of Pop III stars should be hosted by halos with a number density $f_{\rm duty}^{-1}$ times higher. This corresponds to a halo of mass $M_h = 9 \times 10^7 \,\Msun$ or a virial temperature of $5\times 10^4\rm K$, characteristic of an atomic cooling halo. The gas content in such halos is $M_g = (\Omega_b/\Omega_m) M_h = 1.4 \times 10^7 \,\Msun \approx  M_{\star,\rm III}$. Thus, it would be necessary to convert 60\% of the halo gas content into $m_\star = 1000\ \Msun$ stars in $\ll 1.9\ \rm Myr$ to produce the required luminosity. Hence, if possible at all, very extreme conditions would be required by the Pop III interpretation.
Such a high star formation efficiency should also be a relatively rare phenomenon, as implied by the Thompson optical depth of the CMB \citep{visbal:2015} and by the impact on the IGM temperature of binary BH remnants \citep{mirabel:2011}.

Although extreme, the Pop III scenario cannot yet be completely excluded and should be considered in parallel to the PBH solution proposed here. The HeII$\lambda1640$ line luminosity produced by a $1000\ \Msun$ Pop III star is $4.2 \times 10^5\ \Lsun$. The total luminosity of the cluster is then  $L_{\rm HeII} = 3.4 \times 10^9\ \Lsun$. This corresponds to a line flux $\approx 2.9 \times 10^{-18} \,\rm erg\, s^{-1}\,cm^{-2}$ ($\approx 1.3 \times 10^{-19} \,\rm erg\, s^{-1}\,cm^{-2}$) at $z=17$ ($z=25$), i.e., it should be detectable by the NIRSpec deep pointings of a JADES-like survey \citep{eisenstein:2023}.
Interestingly, the line may contaminate the F277W (F444W) filter band for $z=17$ ($z=25$) sources, making their spectra bluer (redder). We caution that HeII$\lambda1640$ could also be excited by PBH emission.

We also consider that before such massive stars collapse into a stellar BH at the end of their lives, they may release non-negligible amounts of heavy elements  (primarily CNO) into the surrounding medium via mass-loss. According to \citet{Marigo03}, a rotating $10^3\,\Msun$ Pop III star produces $(1.18, 1.7\times 10^{-5}, 1.5)\, \Msun$ of (C, N, O), respectively. Hence, once diluted in a gas mass approximately equal to $M_{\star,\rm III}$, this corresponds to a gas metallicity $Z \approx 0.11\,\Zsun$. These elements could potentially generate UV lines detectable by JWST (CIV$\lambda1549$, CIII]$\lambda1907,1909$) or by ALMA (e.g. [OI]63$\mu$m, [OIII]88$\mu$m), thereby enabling discrimination between the PBH and Pop III scenarios. 

Perhaps the most promising perspective to clarify the nature of these sources is to use gravitational waves (GW). Stellar BHs left by $m_\star= 10^3\ \msun$ Pop III stars could merge and produce a GW signal. Neglecting mass loss and assuming complete collapse, the chirp mass of the binary is ${\cal M} = 870\, \Msun$, and the signal would peak at a characteristic frequency $f_m \approx 10$ Hz. The strain is then $h \approx 10^{-21} ({\rm Gpc}/D_L)$, where the luminosity distance to $z=25$ is $D_L=295\ \rm Gpc$, yielding $h \approx 3.4\times 10^{-24}$. This signal corresponds to a signal-to-noise ratio S/N = 3.4 for the expected sensitivity of the Einstein Telescope at 10 Hz \citep{ET25}. Clearly, this is an intriguing possibility that deserves further study. 

\section{Summary}

We have shown that the discovery of nine candidate galaxies at $z=17$ and $z=25$, if confirmed, is virtually impossible to reconcile with current galaxy formation model predictions. However, the implied UV luminosity density can be produced by a relatively small ($\fpbh\approx 10^{-8}$) population of primordial black holes (PBHs) described with a lognormal distribution of characteristic mass $\Mpbh = 10^{4-5} \,\Msun$, typically residing in low-mass halos ($M_h \approx 10^{7} \, \Msun$), and accreting at a moderate fraction of the Eddington luminosity, $\lambda_E \simeq 0.36$. Both the point-like morphology and the observed blue spectra of the detected sources are consistent with this hypothesis, which also complies with current constraints on the mass and abundance of PBHs.

PBH sources precede the first significant episodes of cosmic star formation. At later times, as star formation initiates, PBH emission becomes comparable to or subdominant with respect to that of the hosting galaxy. If ultra-early sources are powered purely by accretion, this strongly disfavors seed production mechanisms requiring the pre-existence of stars (massive or Pop III stars or clusters) or their UV radiation (direct collapse BHs), leaving PBHs as the only alternative solution currently available.

Alternative explanations, such as isolated large clusters ($\approx 10^7 \,\Msun$) of massive ($m_\star=10^3 \,\Msun$) Pop III stars, are only marginally viable. They require extreme and very unlikely conditions that can nevertheless be tested via UV (e.g. HeII, CIII]) and FIR (e.g. [CII], [OIII]) emission lines or gravitational waves.

\begin{acknowledgements}
We thank S. Carniani, M. Castellano, E. Ntormousi, E. Parlanti, P. P\'erez-Gonz\'alez, G. Rodighiero, P. Rosati, S. Salvadori, L. Vallini, E. Vanzella, S. Vergani, A. Zanella, and F. Ziparo for useful discussions.
AF acknowledges support from the ERC Advanced Grant INTERSTELLAR H2020/740120.
Partial support (AF) from the Carl Friedrich von Siemens-Forschungspreis der Alexander von Humboldt-Stiftung Research Award is kindly acknowledged.
We gratefully acknowledge the computational resources of the Center for High Performance Computing (CHPC) at SNS.
We acknowledge usage of \code{Wolfram|Alpha}, the \code{Python} programming language \citep{python2,python3}, \code{Astropy} \citep{astropy}, \code{corner} \citep{corner},  \code{emcee} \citep{emcee}, \code{hmf} \citep{murray:2013}, \code{Matplotlib} \citep{matplotlib}, \code{NumPy} \citep{numpy}, \code{SciPy} \citep{scipy}, \code{multiprocess} \citep{mckerns:2012, pathos}, and statistical approximations from \citet{romeo:2003}.
\end{acknowledgements}

\bibliographystyle{aa_url}
\bibliography{aa54728-25}

\end{document}